\documentclass{iau_no_copyright} 
\usepackage{graphicx}
\usepackage{natbib}
\usepackage{aas_macros}

\title[3D Modeling of the Structure and Dynamics of a Main-Sequence F-type Star] 
{3D Modeling of the Structure and Dynamics of a Main-Sequence F-type Star}

\author[Irina N. Kitiashvili \& Alan A. Wray]   
{Irina N. Kitiashvili$^1$ \and Alan A. Wray$^2$}

\affiliation{$^1$NASA Ames Research Center\\ 
	Moffett Field, MS 258-6, Mountain View, USA \\ email: {\tt irina.n.kitiashvili@nasa.gov}\\[\affilskip]
	$^2$NASA Ames Research Center\\ Moffett Field, MS 258-6,
	Mountain View, USA \\ email: {\tt alan.a.wray@nasa.gov}}

\pubyear{2020}
\volume{354}  
\jname{Solar and Stellar Magnetic Fields: Origins and Manifestations}
\editors{A.G. Kosovichev, K. Strassmeier \& M. Jardine, eds.}

\begin{document}
\maketitle

\begin{abstract}
Current state-of-the-art computational modeling makes it possible to build realistic models of stellar convection zones and atmospheres that take into account chemical composition, radiative effects, ionization, and turbulence. The standard 1D mixing-length-based evolutionary models are not able to capture many physical processes of the stellar interior dynamics. Mixing-length models provide an initial approximation of stellar structure that can be used to initialize 3D radiative hydrodynamics simulations which include realistic modeling of turbulence, radiation, and other phenomena.
 
In this paper, we present 3D radiative hydrodynamic simulations of an F-type main-sequence star with 1.47 solar mass. The computational domain includes the upper layers of the radiation zone, the entire convection zone, and the photosphere. The effects of stellar rotation is modeled in the f-plane approximation.
These simulations provide new insight into the properties of the convective overshoot region, the dynamics of the near-surface, highly turbulent layer, and the structure and dynamics of granulation. They reveal anti solar-type differential rotation and latitudinal dependence of the tachocline location.
\end{abstract}

\section{Introduction}
A dramatic increase in observational data from NASA's Kepler, K2, and TESS missions and supporting ground-based observatories has opened new opportunities to investigate the internal structure, dynamics, and evolution of stars and their atmospheres. However, analysis and interpretation of the observations are challenging, especially for characterizing the structure and dynamics of the outer stellar convection layers and atmospheres. Until recently, only models based on mixing-length theory (MLT) were used to investigate the interior structure and oscillations.

Numerous observed phenomena such as superflares, activity cycles, starspots, and oscillation properties all reflect the properties and dynamics of turbulent convection in the outer layers of stars \citep[e.g.][]{Carroll2014,Balona2015a, Antoci2019,Mathur2019,Strassmeier2019}. Rapidly growing computational capabilities have enabled 3D stellar hydrodynamic simulations on local scales \citep[e.g.][]{Beeck2015,Kitiashvili2012,Kitiashvili2016,Salhab2018}.

In this paper, we present results of 3D radiative hydrodynamic simulations of an F-type main-sequence star of 1.47M$_\odot$, with and without rotation. We discuss the physical properties of the stellar convection and its interactions with the radiative zone in the overshoot layer.

\begin{figure}[b]
	\begin{center}
		\includegraphics[width=5.2in]{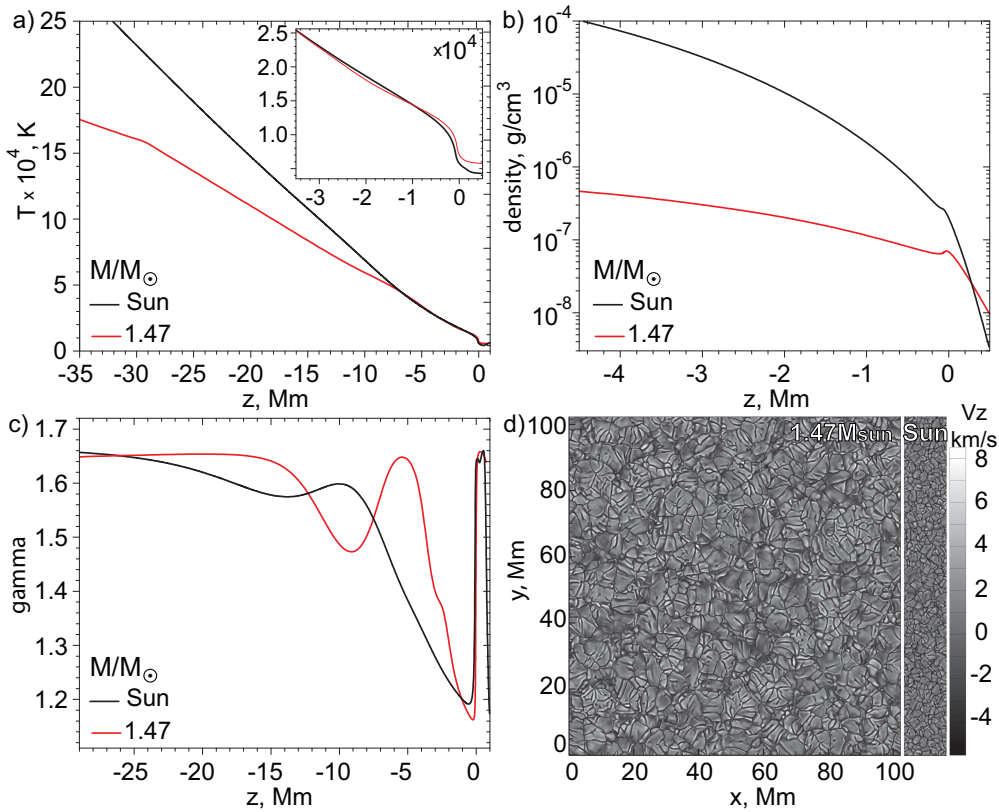} 
		\caption{The mixing-length 1D model of the 1.47~M$_\odot$ F-type star used as the initial conditions to perform the 3D radiative realistic modeling (red curves): a) temperature, b) density, and c) gamma. The black curves show the corresponding mixing-length model of the Sun for comparison. Panel d) shows a snapshot of the vertical velocity at the stellar photosphere. A vertical strip on the right shows the simulations of the solar surface for comparison of the surface granulation scales.}
		\label{model_init}
	\end{center}
\end{figure}

\section{Numerical setup}
The modeling is performed using the StellarBox code \citep{Wray2018}. This code performs modeling from first principles, taking into account the realistic chemical composition, equation of state, radiative transfer, and the effects of turbulence. Starting from a 1D mixing-length model of the internal structure (Fig.~\ref{model_init}a-c) obtained with CESAM code \citep{Morel1997,Morel2008}, we produced a series of 3D dynamical simulations of an F-spectral type main-sequence star of 1.47M$_\odot$ both with rotation and without rotation. The computational domain covers about 20\% of the stellar radius (or 51~Mm in depth), which includes the entire convection zone and the upper layers of the radiative zone. The simulations include a 1~Mm high atmospheric layer. The horizontal size of the computational domain is 102.4~Mm with a resolution of 100~km; the vertical resolution increases with depth from 25~km at the atmosphere to 183~km near the bottom boundary. In the StellarBox code, stellar rotation is implemented using the f-plane approximation.

\section{Thermodynamic structure of non-rotating stars with shallow convection zones}
The turbulent dynamics of stars with relatively shallow convection zones attracts our interest because it allows us to investigate the properties of turbulent convection across the whole convection zone and the transition into the convectively stable radiative zone with a high degree of realism. 

\begin{figure}[b]
	\begin{center}
		\includegraphics[width=5.2in]{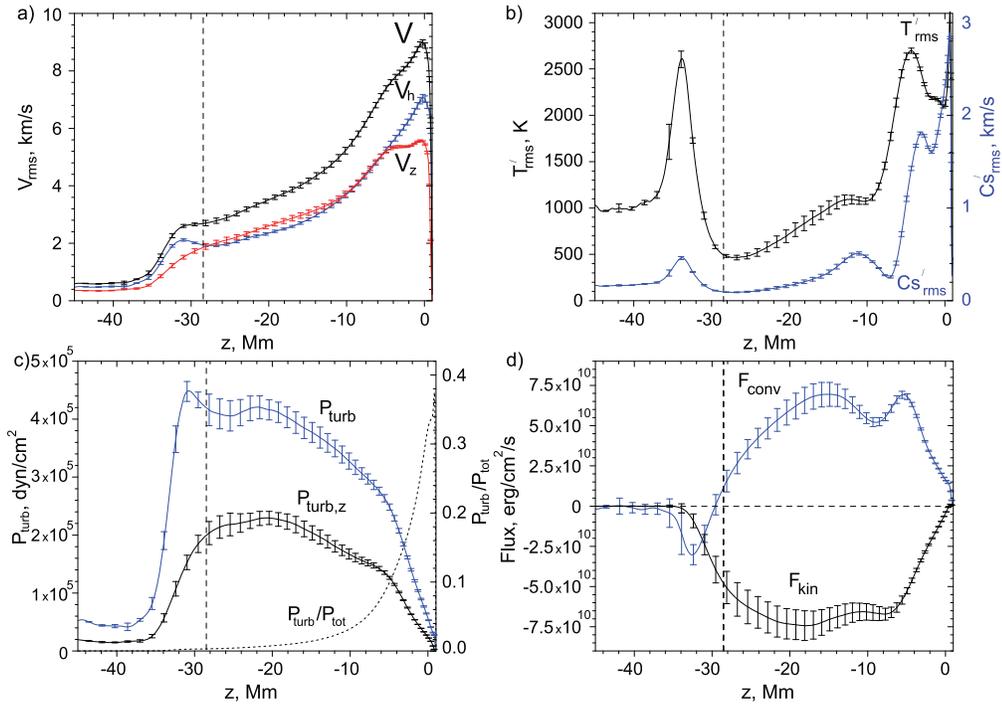} 
		\caption{Vertical profiles, obtained from the 3D numerical radiative hydrodynamic simulation of a 1.47M$_{\odot}$ star: (a) rms of velocity V (black), vertical Vz (red) and horizontal Vh (blue) components of velocity; (b) rms of temperature $T'$ (black) and sound-speed $c_{S'}$  (blue) perturbations; (c) turbulent pressure $P_{turb}$ (blue curve), turbulent pressure of the vertical motions $P_{turb,z}$ (black), and the ratio of turbulent pressure $P_{turb}$ to total pressure $P_{turb}+P$; and (d) convective energy flux $F_{conv}$ (blue curve) and kinetic energy flux $F_{kin}$ (black), calculated according to the formulation of \cite{Nordlund2001}. Vertical dashed lines indicate the bottom boundary of the convection zone of the corresponding 1D stellar model: z$_{cz}=-28.5$~Mm.}
		\label{1Dprofiles}
	\end{center}
\end{figure}

Our simulation results have revealed that a characteristic feature of the stellar photosphere of F-type stars is the co-existence of several granulation scales \citep{Kitiashvili2016}. The convection pattern represents well-defined `small’ and `large’ granules self-organized into clusters (Fig.~\ref{model_init}d). In deeper layers of the convection zone, the scale of the convection pattern gradually increases. In the intergranular lanes, some downdrafts penetrate through the entire convection zone, reaching velocities of more than 20~km/s. These downdrafts penetrate into the radiative zone, form an overshoot layer and cause local heating (Fig.~\ref{1Dprofiles}b). This increases the local sound speed and initiates density fluctuations that are a source of internal gravity oscillations ($g$-modes). 

It is informative to consider mean vertical profiles of the fluctuations for various turbulence quantities and the convective and kinetic energy fluxes from the stellar photosphere down to the radiative zone (Fig.~\ref{1Dprofiles}). The strength of the horizontal flows gradually decreases from the photosphere to the bottom of the convection zone. In the overshoot region, the amplitude of the horizontal velocities increases slightly due to the splashing of downdrafts (Fig.~\ref{1Dprofiles}a). The rms vertical velocity profile shows a strong deviation from the horizontal flows near the surface layer. In particular, a broad bump in horizontal velocity located from 0 -- 5~Mm below the surface corresponds to one of the characteristic scales of the granulation layer. Below 10 Mm, the fluctuations of vertical velocity are stronger than the horizontal ones, probably due to the penetrating high-speed downdrafts. In the overshoot layer, the velocity strength sharply decreases.

\begin{figure}[b]
	\begin{center}
		\includegraphics[width=5.2in]{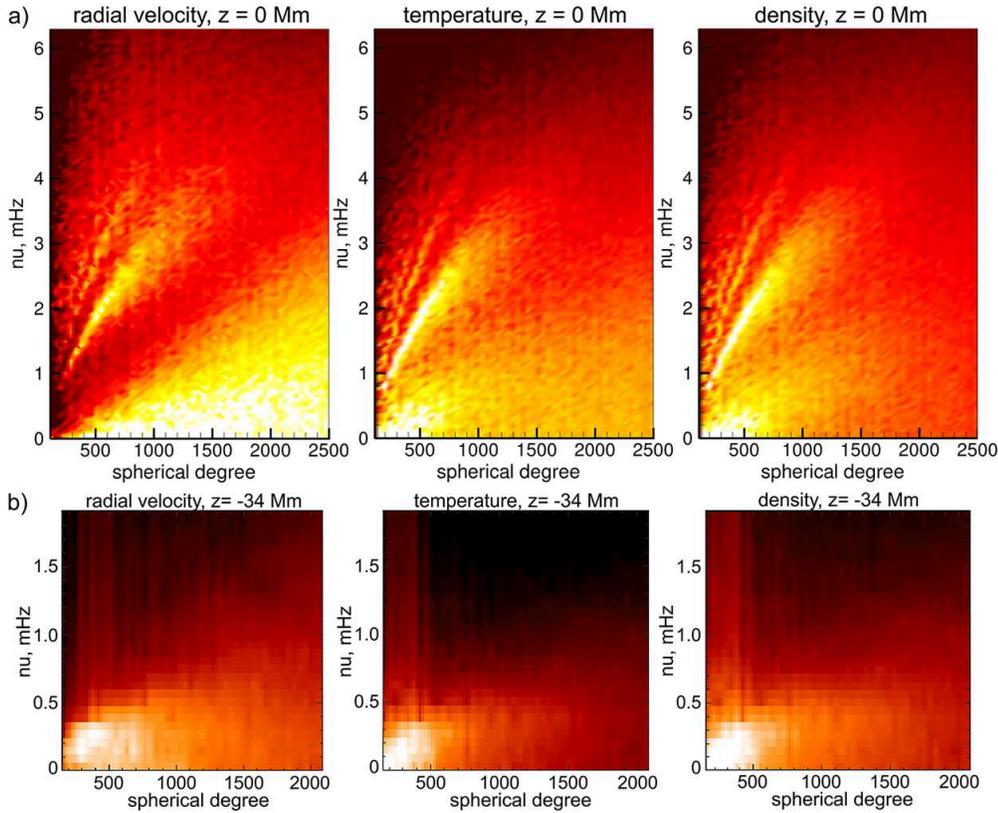} 
		\caption{Power spectra of the radial velocity, temperature, and density, represented as functions of angular spherical harmonic degree and frequency 		
			for: a) the stellar surface and b) the radiative zone just below the tachocline layer at depth 34~Mm below the photosphere.}
		\label{k-w}
	\end{center}
\end{figure}

The distribution of temperature fluctuations with depth (Fig.~\ref{1Dprofiles}b, black curve) shows significant variations near the surface layers and in the overshoot region. The near-surface fluctuations (peaked at near --5~Mm) are mostly related to strong radiative cooling in the intergranular lanes and are associated with downflows that determine the granulation scales. A sharp increase in temperature fluctuations from --10 to --5Mm corresponds to the He II ionization zone. In this region, the rms of vertical velocity and the convection energy flux (Fig.~\ref{1Dprofiles}a,c) also increase, indicating an enhancement of turbulent convection in the ionization zone. Temperature fluctuations at $z\approx -34$Mm, which is associated with the overshoot region, reflect the intense local heating in these layers due to plasma compression by downdrafts that penetrate through the convection zone.
The rms sound-speed perturbations (Fig.~\ref{1Dprofiles}b, blue curve) have a similar shape, but their amplitude in the overshoot region is much smaller than in the subsurface layers.

The stellar oscillations excited in the convection zone are shown in the form of the power spectra ($l-\nu$ diagrams) near the photosphere (Fig.~\ref{k-w}a) and the overshoot region (Fig.~\ref{k-w}b). At the stellar surface, the power spectra of the radial velocity, temperature, and density clearly show ridges, which correspond to surface gravity $f$-modes and acoustic p-modes. An additional broad ridge visible in the density power spectrum below 1mHz possibly corresponds to internal gravity $g$-modes. However, the power spectra of radial velocity and temperature do not show this ridge. The power spectra of the upper layers of the radiative zone, near the overshoot region, do not show any acoustic-type oscillations (Fig.~\ref{k-w}b), probably because of the flow amplitude. However, all quantities show a signature of $g$-modes, excited in the overshoot region. This may explain why $g$-mode oscillations are not observed even for stars with shallow convection zones, where the surface convective turbulent flows are strong.

\begin{figure}[b]
	\begin{center}
		\includegraphics[width=5.2in]{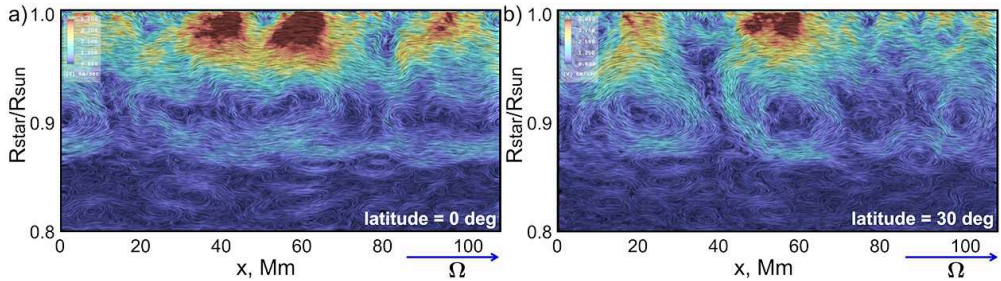} 
		\caption{Comparison of the rotating convection zone with period of rotation of 1 day at different latitudes: a) $0^{\circ}$ (equator), and b) $30^{\circ}$. The dynamical structure of the convective patterns is visualized by the particle tracing method.}
		\label{M147rot1}
	\end{center}
\end{figure}

\begin{figure}[t]
	\begin{center}
		\includegraphics[width=5.2in]{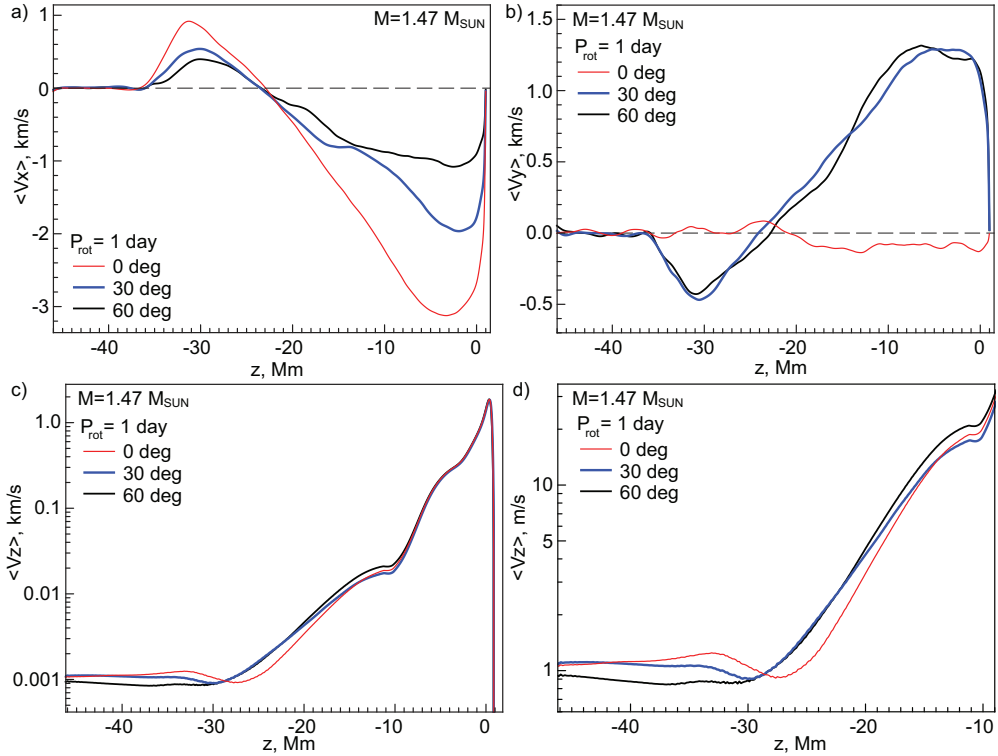} 
		\caption{Effect of stellar rotation on the mean velocity profiles for a rotation period of 1 day at three latitudes: $0^{\circ}$ (red curves), $30^{\circ}$ (blue curves), and $60^{\circ}$ (black curves). Panel a) shows radial profiles of the azimuthal velocity, representing differential rotation. Panel b): Radial profiles of the mean velocity field corresponding to the meridional circulation at different latitudes. Panels c) and d) show the mean radial profiles for the vertical velocity component across the whole domain (panel c), and a zoom into the tachocline layer, d). All profiles are averages in horizontal planes over one stellar hour.}
		\label{M147diff_rot}
	\end{center}
\end{figure}

\section{Interior dynamics of rotating stars} 
Links between stellar rotation and stellar thermodynamic properties are a hot topic in various fields \citep[e.g.][]{Law1981,MacGregor2007,Eggenberger2010,Brito2019}. Therefore, understanding how stellar internal dynamics and structure depends on the rotation period is critical for interpreting observational data.
To study rotation effects, we performed simulations for rotational periods of 1 and 14 days using the f-plane approximation, in which the Coriolis force is assumed constant in horizontal planes. This approximation is applicable for local simulations because the simulation domain covers only $\sim 5.6^{\circ}$ in latitude. The full radiative hydrodynamic simulations are performed for three latitudes: $0^{\circ}$ (equator), $30^{\circ}$, and $60^{\circ}$ for both rotation rates.

\begin{figure}[t]
	\begin{center}
		\includegraphics[width=5.2in]{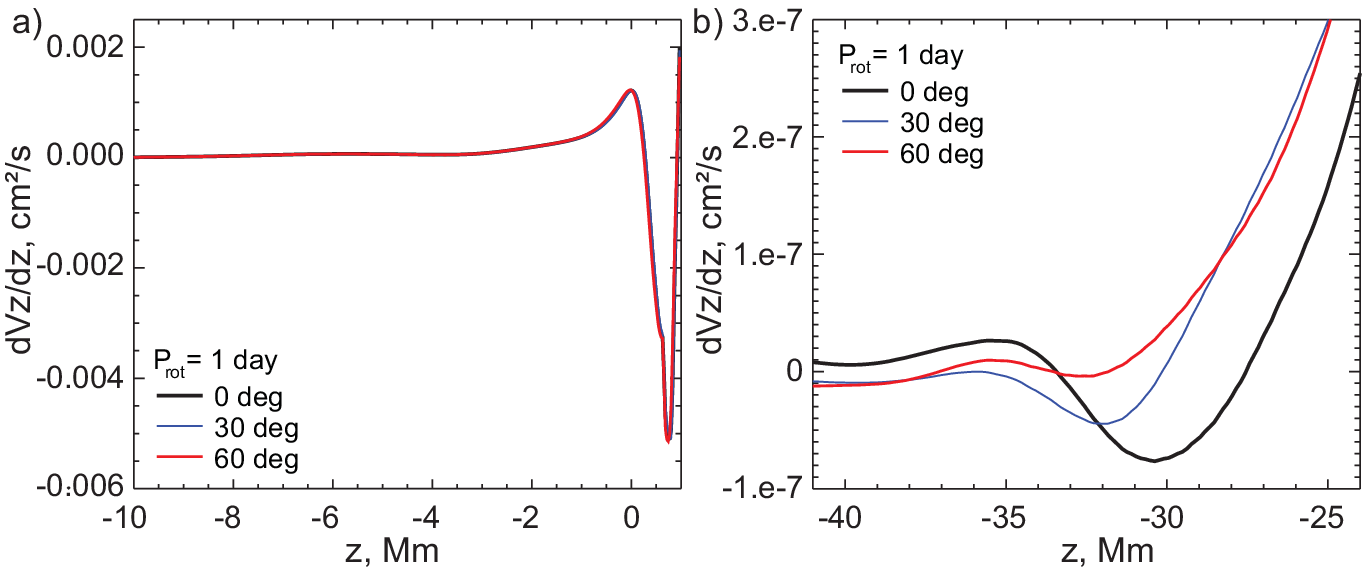} 
		\caption{Radial profiles of the mean vertical velocity time-derivative, $dVz/dt$, for the whole computational domain (panel a) and near the tachocline layer. The profiles were obtained for a rotating main-sequence 1.47~M$_\odot$ star with a period of 1~day at 3 latitudes: $0^{\circ}$ (thick black lines), $30^{\circ}$ (thin blue lines), and $60^{\circ}$ (red curves).}
		\label{M147rot_derivVz}
	\end{center}
\end{figure}

The simulation results show that stellar rotation does not affect the structure and dynamics of the surface layers. However, in deeper layers of the convection zone, two distinct layers can be identified (Fig.~\ref{M147rot1}). 
In the outer layer of the convection zone, large-scale flow patterns are transported in the direction opposite to stellar rotation. Such a distinct structure of the flow in the upper layers of the convection zone represents a subsurface shear flow, similar to the one observed in the Sun. Profiles of the azimuthal velocity, meridional circulation, and the mean vertical velocity (Fig.~\ref{M147diff_rot}a-c) show that the subsurface shear layer extends up to 10.5~Mm into the convection zone.  
The inner layers of the convection zone reveal cellular or roll-like circular motions. The structure and scale of these large-scale eddies depend on latitude. At the equator (Fig.~\ref{M147rot1}a), they are very elongated and often can split into smaller and weaker patterns. At higher latitudes, the eddies are more compact and stable with sizes of about 25~Mm at $30^{\circ}$ latitude for $P_{rot}=1$day (Fig.~\ref{M147rot1}b).

The ability to model the effects of stellar rotation at different latitudes allows us to study the properties of differential rotation and meridional circulation for various periods of rotation and latitudes (Fig.~\ref{M147diff_rot}a,b). The differential rotation shows faster rotation at the equator and slower at $60^{\circ}$ latitude, indicating anti solar type rotation (Fig.~\ref{M147diff_rot}a). In the tachocline layer, reverse azimuthal flows (corresponding to deceleration of the stellar rotation) are present. The relative strength of the return flows correlates with subsurface shear flows at different latitudes. Thus, the reverse flows are weaker at higher latitude, and the subsurface shear flows are also weaker in comparison with the flows at the equator. The strongest subsurface and reverse flow location depends on the latitude. In particular, the maximum of the subsurface shear flow is located closer to the stellar photosphere in comparison with the flows at the equator. The maximum of the reverse flows is shifted into the deeper layers of the stellar interiors in comparison with the flows at higher latitudes. In the case of slower rotation, $P_{rot}=14$~days, these trends also take place but are weaker and develop slowly, requiring a significantly longer simulation run.

Another component of the large-scale flows, perpendicular to the stellar rotation, can be interpreted as meridional circulation. Because of the periodic lateral boundary conditions, the resulting models are able to reproduce only a narrow range of latitudes of the meridional circulations. Figure~\ref{M147diff_rot}b shows the radial profiles of the meridional circulation at the equator,  $30^{\circ}$, and $60^{\circ}$ latitudes. The meridional component of the velocity at the equator fluctuates around zero, meaning that there is no significant mean cross-equator meridional flow, which is not surprising. A weak negative flow velocity in the convection zone is likely due to only 1-hour averaging, but potentially may reflect the flow variations on longer time-scales. This requires additional study. The meridional circulation at $30^{\circ}$ and  $60^{\circ}$ latitudes reveals a consistent meridional circulation structure in the flow variations.

The overall mean distribution of the vertical flows is shown in Figure~\ref{M147diff_rot}c. The vertical flow distribution is the same in the layers from the photosphere down to 9.5Mm below the photosphere. In the deeper layers, the vertical velocities start to deviate at different latitudes (Fig.~\ref{M147diff_rot}c,d). The fastest decrease of the velocity is at the equator, where it reaches a local minimum around 27~Mm, which corresponds to the top of the tachocline layer. After that, the vertical velocity increases until the bottom of the tachocline layer. In the radiative zone, the velocities slowly decrease with depth. Similar behavior of the vertical flows takes place at $30^{\circ}$ latitude, with a minimum at a depth of 30~ Mm below the surface. At $60^{\circ}$, there is no prominent minimum. The vertical velocity derivative reveals a clearer picture (Fig.~\ref{M147rot_derivVz}). In particular, it shows the velocity variations near the stellar surface and in deeper layers. Near the tachocline layer, the derivative minimum at the equator is located at 30~Mm below the stellar surface, at 32~Mm for $30^{\circ}$, and 32.5~Mm for $60^{\circ}$ latitude. This provides evidence that the depth of the tachocline varies with latitude.

\section{Conclusion}
Despite the availability of advanced observational data from modern space and ground instruments, investigation of the dynamics and structure of the surface and subsurface layers of stars is quite challenging. We performed a series 3D radiative hydrodynamic simulations of an F-type star with mass 1.47~M$_\odot$, in which the whole convection zone and upper layers of the radiative zone were included in the computational domain. 
The simulation results reveal the formation of an overshoot layer and also multi-scale populations and clustering of the surface granulation. High-speed convective downdrafts of 20 -- 25km/s penetrate through the convection zone, form an overshoot layer, and contribute to excitation of internal gravity waves ($g$-modes). These waves are identified near the overshoot layer. At the stellar photosphere, these modes are hidden among strong turbulent convective flows, and only $f$- and $p$-modes are clearly displayed in the simulated power spectra. 

Simulating of effects of stellar rotation, for rotational periods of 1 and 14 days at different latitudes, allowed us to identify the formation of a subsurface shear flow and roll-like convective patterns in the deep layers of the convection zone. The radial profiles of the differential rotation indicate that it is of anti solar type. The subsurface shear flow velocity peaks closer to the photosphere at higher latitudes. The meridional circulation profiles do not show a significant difference at $30^{\circ}$ and $60^{\circ}$ latitudes. The simulation results show that the tachocline layer is located deeper and is less prominent at higher latitudes.

Our future plans include expansion of the computational domain higher into the stellar atmosphere and taking into account magnetic fields, as well as developing these types of numerical models for more massive stars.

{\bf Acknowledgment.} The work is supported by NASA Astrophysics Theory Program.

\bibliographystyle{aa}


\end{document}